\documentclass[11pt, nofootinbib]{article}
\usepackage{geometry} 

\usepackage{epsfig}
\usepackage{epstopdf}
\usepackage{amsmath}
\usepackage{bm}
\usepackage{braket}

\newcommand{\thickbar}[1]{\mathbf{\bar{\text{$#1$}}}}  
  
\begin{document}

\title{Are Retrocausal Accounts of Entanglement \\ Unnaturally Fine-Tuned?}
\author{D. Almada, K. Ch'ng, S. Kintner, B. Morrison and K.B. Wharton\thanks{San Jos\'e State University, Department of Physics and Astronomy, San Jos\'e, CA 95192-0106} }

\date{} 


\maketitle

\abstract{An explicit retrocausal model is used to analyze the general Wood-Spekkens argument \cite{WS} that any causal explanation of Bell-inequality violations must be unnaturally fine-tuned to avoid signaling.  The no-signaling aspects of the model turn out to be robust under variation of the only free parameter, even as the probabilities deviate from standard quantum theory.  The ultimate reason for this robustness is then traced to a symmetry assumed by the original model.  A broader conclusion is that symmetry-based restrictions seem a natural and acceptable form of fine-tuning, not an unnatural model-rigging.  And if the Wood-Spekkens argument is indicating the presence of hidden symmetries, this might even be interpreted as supporting time-symmetric retrocausal models. }

\section{Introduction}

Bell's theorem has ruled out local past-common-cause explanations of some correlations observed in entanglement experiments, but this has not stopped research into more general causal explanations of such phenomena in terms of spacetime-local beables.  The options on the table include superluminal causal influences, retrocausal explanations, and a casual restriction on the measurement settings themselves.  It has recently been noted by Wood and Spekkens that all of these causal explanations would seem to require substantial ``fine-tuning'' in order to prevent the possibility of nonlocal signaling. \cite{WS}

Indeed, our experience and intuition tell us that causal relationships almost always include a signal channel.  If Alice can cause something near Bob, we would naturally expect that she could signal to Bob as well. Wood and Spekkens correctly point out that the no-signaling theorems (combined with Bell-inequality violations) are therefore a major challenge to \textit{any} causal explanation of entanglement phenomena.

However, our intuition on matters of causation and signaling have arisen from common experience.  We are used to having access to certain resources (low entropy sources) and we are used to being subject to certain constraints (the second law of thermodynamics).  In different situations, our intuitions are not so accurate.  For example, in a maximum-entropy universe, filled with a blackbody electromagnetic spectrum, it would be impossible to use a polarizing cube to signal. (This assumes the cube was also at thermal equilibrium with the field, preventing it from casting shadows.)  

Certainly, our choice of cube-orientation could \textit{cause} an outcome; given the input radiation to the cube, different choices would correspond to different output radiation.  But without knowing the microscopic fluctuations in the incoming radiation, there would be no way to send a signal.  Indeed any resulting signal would be a provable failure of the second law, with the cube's controller playing the role of Maxwell's Demon.  So in this (entropically highly probable) situation we would \textit{expect} generic causation without signaling. 

With this point in mind, one strategy in addressing the Wood-Spekkens challenge would be to find a causal explanation in a framework where our intuition does not have much experience.  The obvious candidate from the above list of possibilities is retrocausal accounts of entanglement (especially given the possible relevance of the thermodynamic arrow of time, highlighted by the above example and related discussion \cite{IJQF}).  Another strategy would be to find a specific model that could be analyzed on charges of ``fine-tuning''; specific models are certainly more amenable to such analysis than generic arguments.  This paper therefore takes up the Wood-Spekkens challenge by analyzing a recent quantitative retrocausal account of Bell-inequality violations.

Although retrocausal models may be instinctively distasteful, the alternative to taking up their challenge is to retreat from causal explanations of these phenomena entirely (dropping back to mere inferential explanations).  As Wood and Spekkens note, ``the idea of explaining correlations \textit{causally} appears to us to be central to the scientific enterprise." (\cite{WS}, emphasis added)  We fully agree with this point; no matter how distasteful retrocausality may appear, giving up on any attempt to find a causal explanation of reproducible phenomena would be far worse.  

After outlining the retrocausal model in Section 2, the Section 3 analysis will consider variations to a natural free parameter in the model, a parameter that must be very small to recover the standard quantum probabilities.  We will show that varying this parameter deviates the results from quantum mechanics, but still does not allow any signaling.  This curious result therefore implies a large class of causal-but-non-signaling models.  According to Wood-Spekkens, \textit{all} of these models are likely fine-tuned in some manner.  (Another implication is that these models allow a smooth deviation from the Tsirelson bound, which may be useful in probing this bound's ultimate nature.)

The section 4 analysis then finds the source of the fine-tuning in these explicit models, which can be traced to a basic symmetry.  In general, imposing symmetries is itself a sort of ``fine tuning'', in that a large parameter space is restricted to a special (symmetrical) subset.  Indeed, ``symmetry protection'' is a standard argument when justifying fine-tuned masses in particle physics. \cite{LHC}  The question of whether such restrictions are \textit{unnaturally} fine-tuned, then, turns on the naturalness of the symmetry in question.  In section 5, this question will be explored in a more general context, considering all live options for causal explanations of entanglement-based correlations.

\section{The Retrocausal Model}

\subsection{Schulman's One-Particle Model}

The two-particle entanglement model we shall be considering has been outlined in a recent paper \cite{Wharton14}, with further motivation to be found elsewhere \cite{WMP,LOQT}.  It is based on a one-particle model proposed by Schulman \cite{Schulman}, and understanding this one-particle model is central to understanding the two-particle model.  Indeed, the two-particle model is essentially a trivial extension.

Schulman's model applies to a single spin-1/2 particle on a known path, subject to two consecutive measurements.  The spin-vector $\bm{S}$ represents the actual state of the system at any given time (say, a vector on the Bloch sphere).  This vector is meant to map to the expectation value of the spin operator $\langle \bm{S}\rangle$ for a pure state, at least at points where the state is explicitly measured.  Schulman proposed an \textit{ansatz} that between any two measurements, this spin-vector is permitted to undergo an anomalous rotation through a net angle $\alpha$ with a global probability proportional to 
\begin{equation}
\label{eq:Wdef}
W(\alpha)=\frac{1}{\alpha^2+\gamma^2}.
\end{equation}
Here $\gamma$ is a small free parameter in the model that cannot go to zero without making $W(\alpha)$ ill-defined for anomaly-free histories.  It is also important to note that these (unnormalized) global probability weights are assigned to entire histories, not particular instants.  Finally, note that these rotation anomalies are confined to a single plane of rotation, as determined by the boundary conditions, which will now be discussed.

All spin measurements on the particle are imposed as boundary conditions.  This is considered normal for the preparation; if one initially measures the spin in some particular direction $\bm{S_i}$, it is standard to assume that the prepared spin-vector is aligned with $\bm{S_i}$.  But here, there is no projection postulate upon measurement; if the second measurement finds the spin to be in some particular direction $\bm{S_f}$, then this direction is imposed as a \textit{final boundary condition} on the spin-vector, just like the preparation.  If $\bm{S_i}\ne\bm{S_f}$, and if there is no standard dynamical process which would take the former to the latter, then an anomalous rotation must necessarily occur.  This anomaly rotates the spin-vector by an angle $\alpha$, with a probability distribution proportional to $W(\alpha)$ over all allowable rotations.

The only asymmetry between past and future in Schulman's model comes in as a restriction on the experimenters themselves.  If Alice is making the first measurement, and Bob is making the second measurement, both Alice and Bob can arbitrarily choose the angles of their spin-measurement settings.  (Their choice arises from outside the system of interest, effectively as free external boundary conditions on the particle.)  But only Alice can select the \textit{actual} outcome (choosing one of two possibilities, $\pm\hbar/2$), and send the corresponding particle to Bob.  Bob has no control over his actual outcome, and can only read off the sign of the result. 

We take this obvious and well-known asymmetry to be a consequence of the second law of thermodynamics constraining Alice and Bob.  For whatever reason, they only have knowledge of their past, not their future.  So Alice can block transmissions that do not have the outcome she desires, but by the time Bob learns his outcome, it is too late to block the transmission at his end.  (More accurately, at the spacetime locations where Bob thinks he could choose to block a transmission, he does not have epistemic access to the outcome of his measurement.)

Suppose Alice and Bob make consecutive spin measurements, each in some particular chosen direction, and their two chosen angles differ by $\theta$.  Alice chooses her spin-vector output to be aligned with her setting angle; her outcome is always $+\hbar/2$.  But Bob cannot make such a choice; the spin-vector will either be aligned ($+\hbar/2$) or anti-aligned ($-\hbar/2$) with his chosen direction.  These two possibilities are the two possible final boundary conditions on the particle.  Therefore, the net anomalous rotation between measurements must be either $\theta$ (mod $2\pi$) or $\pi+\theta$ (mod $2\pi$), corresponding to Bob's two possible outcomes.  

But note that $W(\alpha)$ is not a cyclic function; $W(\theta)$ is different from $W(\theta+2\pi)$.  To generate probabilities, then, one must sum over all possible anomalous rotations that lead to the same result.  Schulman finds this ratio of probabilities as
\begin{equation}
\label{eq:Sch1}
\frac{P(\theta)}{P(\pi+\theta)}=\frac{ \displaystyle\sum_{n=-\infty}^{\infty}W(2n\pi+\theta)}{\displaystyle\sum_{n=-\infty}^{\infty}W(2n\pi+\pi+\theta)}=\frac{\cos^2(\theta/2)+\sin^2(\theta/2)\tanh^2(\gamma/2)}{\sin^2(\theta/2)+\cos^2(\theta/2)\tanh^2(\gamma/2)}.
\end{equation}

In the limit that $\gamma\to 0$, this reduces to the standard Born rule probabilities, because the $\tanh^2(\gamma/2)$ terms go to zero.  Specifically, since \textit{one} of these two outcomes must occur, normalization allows one to reduce the ratio of probabilities to absolute probabilities.  In this case, the probability Bob finds the ($+\hbar/2$) outcome is $P(\theta)=\cos^2(\theta/2)$.  And since $\gamma$ is an arbitrary parameter, Schulman's \textit{ansatz} can be made arbitrarily close to the Born rule. 

Note that Schulman's one-particle model is effectively retrocausal.  If Bob chooses the same measurement angle as Alice, there will never be any anomalous rotation. But if Bob chooses a different angle, then \textit{before} he makes his choice, an anomalous rotation may occur. Because Bob's future choice determines whether or not past anomalies will occur, this is a retrocausal model -- as generally is any model with a choosable final boundary condition.  The ``bilking" argument (that Bob may find out the spin-vector and therefore learn about his future choice before he makes it) is naturally resolved by the link between measurements and boundary conditions; the only way to directly learn anything about the spin-vector is to physically make an intermediate measurement, which would change the boundaries of the experiment.  Also note that this block-universe style of retrocausal model does not remove Bob's free choice of setting; that is coming from outside the system of interest, as an effective external boundary.

Even at this stage, Schulman's model is an example of (retro-)causation without {(retro-)}signaling; Alice can't learn about intermediate anomalous rotations, and so can't detect Bob's future setting.  The Wood-Spekkens paper then raises the question of whether this one-particle model is already fine-tuned, even before the extension to entanglement.  For now, we simply note that the below analysis will consider variations to this base model, and therefore will effectively explore this very question.  

\subsection{Extension to an Entangled System}

Schulman's one-particle model does not assign conditional probabilities to instantaneous states, but rather joint probabilities to entire histories: it is the entirety of the anomalous rotation angle $\alpha$ that appears in Eqn (\ref{eq:Wdef}).  This feature allows a natural extension of the one-particle model to a two-particle system, where each particle has a local spin-vector.  For two spin-1/2 particles, if $\bm{S}_1$, the spin-vector of particle $P1$, undergoes an anomalous rotation of angle $\alpha$, and $\bm{S}_2$, the spin-vector of particle $P2$, undergoes an anomalous rotation of angle $\beta$, it is natural to define the unnormalized joint probability of these two events to be simply $W(\alpha) W(\beta)$. 

Remarkably, this usual combination of joint probabilities is all that is needed to recover a retrocausal explanation of Bell-inequality violations, in terms of spacetime-local beables (the two spin-vectors).  Specifically, we are interested in reproducing the Born rule as applied to an entangled singlet (Bell) state $(\ket{01}-\ket{10})\sqrt{2}$, but without using such a non-local state to describe the system.  

Instead, we can simply model the preparation of such an entangled system as a local, classical correlation between two distinct and localized spin-vectors: $\bm{S}_1$ and $\bm{S}_2$ are initially constrained to point in opposite directions, but that direction is completely unspecified.  This is \textit{not} to say that each vector has a uniform probability distribution over all directions; the probabilities are assigned to entire histories, not instantaneous states. Rather, the absolute direction of $\bm{S}_1$ is unconstrained, meaning that all possible directions should be considered when constructing the possibility space on which the weights $W(\alpha) W(\beta)$ can be assigned.  But with this caveat aside, this merely describes ``classical entanglement": a purely conventional correlation between two unknown parameters.  Learning the initial $\bm{S}_1$ would precisely inform us about the initial $\bm{S}_2$, and vice-versa.  

After this classically-entangled pair of particles is produced, $P1$ is sent to Alice, and $P2$ is sent to Bob.  Both Alice and Bob freely choose spin measurements to perform on the particles they receive.  Using the above analysis, this is effectively Schulman's one-particle model applied twice; $\bm{S}_1$ undergoes an anomalous rotation of $\alpha$ to match one of the two allowed outcomes of Alice's setting, and $\bm{S}_2$ undergoes an anomalous rotation of $\beta$ to match one of the two allowed outcomes of Bob's setting.   What links the probabilities of the two outcomes is merely the classical entanglement imposed locally at the preparation.

From (\ref{eq:Wdef}) it follows that $W(0)\gg W(\alpha)$ for any $\alpha\gg\gamma$.  But the free parameter $\gamma$ must be very small, or else (\ref{eq:Sch1}) would deviate from Born rule probabilities \cite{Schulman}.  And if $\gamma$ is below experimental precision, $\alpha$ would be much larger than $\gamma$ for \textit{any} measurable non-zero angle.  The joint probability $W(\alpha)W(\beta)$, then, will be dominated by cases where either $\alpha$ or $\beta$ is zero (or at least comparable to $\gamma$, which is indistinguishable from zero).  In other words, the small value of $\gamma$ makes two separate anomalies very unlikely: the probability distribution $W(\alpha)W(\beta)$ will essentially force all of the anomaly to occur on either $P1$ \textit{or} $P2$, not a combination of both.

An explicit example may be useful to highlight this crucial conclusion.  Suppose Alice and Bob choose setting angles that are separated by an acute angle $\theta$.  In the special case that the original $\bm{S}_1$ (and $-\bm{S}_2$) is aligned halfway between these two settings, both particles need at least a $\theta/2$ anomalous rotation to match the future boundary conditions. Since $\gamma$ is much smaller than $\theta$, the biggest possible value of $W(\alpha)$ is about $4/\theta^2$, and $W(\beta)$ is also maximum at $4/\theta^2$, so the joint probability of this scenario is $W(\alpha)W(\beta)=16/\theta^4$.  There are less likely cases as well, including anomalous rotations of $(\pi - \theta/2)$, etc.

But in the special case that $\bm{S}_1$ is exactly aligned with Alice's future setting, then $W(\alpha)$ is $1/\gamma^2$, and $W(\beta)$ is $1/\theta^2$ (all the anomaly is on $\bm{S}_2$).  The joint probability of this scenario is $W(\alpha)W(\beta)=1/(\theta\gamma)^2$. In the limit that $\gamma\ll\theta$, this joint probability is overwhelmingly more probable than the case in the previous paragraph. So itÕs overwhelmingly more probable that the original spin orientation will happen to be aligned with one of the future settings. (The retrocausality is quite evident here; more on this below.)

The conclusion that either $\alpha$ or $\beta$ must be indistinguishable from zero makes analysis of the general case trivial.  Again, define the net angle $\theta$ between Alice's and Bob's measured spin directions.  If $\alpha=0$ and all of the anomalous rotation is in $\beta$, then this is essentially the same one-particle problem as before:  Alice's setting chooses the axis that the original $\bm{S}_1$ is aligned with, and the preparation ensures that $\bm{S}_2$ is in the opposite direction.  This vector must then anomalously rotate by either $\theta$ (mod $2\pi$) or $\theta+\pi$ (mod $2\pi$) to match Bob's setting, and the probabilities of such a rotation are the same as in the one-particle case.  The same probabilities can also be recovered if $\beta=0$, by simply switching Alice and Bob in this analysis.

The only complication here is that Alice can no longer choose her outcome; she's now in the same position as Bob, at another future boundary.  This means that there are now \textit{four} relevant joint probabilities to consider instead of two.  Alice's measured spin-vector will either be aligned with her measurement setting (call this outcome $A=0$) or anti-aligned ($A=1$).  Similarly, Bob's measured spin-vector will either be aligned with his measurement setting (call this outcome $B=0$) or anti-aligned ($B=1$).  For the two outcome scenarios where $A\ne B$, the required anomalous rotation (on either $\alpha$ or $\beta$, but not both) is simply $\theta$ (mod $2\pi$).  For the two outcome scenarios where $A=B$, the required rotation is $\pi+\theta$ (mod $2\pi$).  But since the anomaly is overwhelmingly likely to be on just one particle (either $\bm{S}_1$ or $\bm{S}_2$), the calculation already performed in (\ref{eq:Sch1}) is also the answer here.  Given the above assumption that $\gamma$ is smaller than any measurable angle,
\begin{equation}
\label{eq:corr}
\frac{P(A\ne B)}{P(A= B)}=\frac{P(\theta)}{P(\pi+\theta)}=\frac{\cos^2(\theta/2)+\sin^2(\theta/2)\tanh^2(\gamma/2)}{\sin^2(\theta/2)+\cos^2(\theta/2)\tanh^2(\gamma/2)}.
\end{equation}
In the $\gamma\to 0$ limit one can normalize these probabilities and find that the correlation $P(A\ne B)-P(A=B)$ becomes simply $\cos(\theta)$, as would be expected from traditional quantum theory on a Bell state.  

This correlation violates the Bell inequalities, but does not violate Bell's theorem because it is retrocausal.  Specifically, Bell assumed that any hidden variable distribution could not depend on the future measurement settings.  To see how this assumption is explicitly violated here, notice that (because $\alpha=0$ or $\beta=0$) the original spin-vectors at preparation will always be aligned (or anti-aligned) with one of the measurement settings eventually chosen by Alice or Bob.  One knows nothing about this ``hidden variable'' until after Alice and Bob make their setting decisions, and as above, it is immune to the bilking argument.  Because the hidden variable of the initial spin-vector alignment is effectively caused by the eventual measurement choice, the premises behind Bell's theorem are explicitly violated; these so-called ``retrocausal loopholes'' are also available for other no-go theorems.

With this model in hand, we can now address the Wood-Spekkens challenge via explicit analysis, and attempt to determine whether or not the no-signaling aspects of this model are fine tuned.

\section{Tuning the Model}

The above model does have one obvious free parameter, $\gamma$, which cannot quite be zero without making the probabilities ill-defined.  This section will explore the consequences of varying $\gamma$ in (\ref{eq:Wdef}).  The Born rule is only recovered in the limit that $\gamma\to 0$, so varying $\gamma$ will deviate from the Born Rule.  If such deviations would also allow signaling between Alice and Bob, this would be evidence for Wood-Spekkens fine-tuning, and indeed would demonstrate the precise parameter in question.  However, it will turn out that this is not the case.

\subsection{Effect on Probabilities}

The above analysis for the entangled state can be reproduced for the case where $\gamma$ is not eventually set to zero.  Again, $\theta$ is the angle between Alice's setting and Bob's setting. Outcome $A$ is when Alice's measured spin vector $\bm{S}_1$ points in the direction of her chosen setting; outcome $\thickbar{A}$ is when $\bm{S}_1$ is opposite this direction.  Similarly, Bob's two outcomes are denoted by $B$ and $\thickbar{B}$.

The first difference, due to a non-zero $\gamma$, is that the entire anomaly is no longer forced to be on one particle or the other.  (That earlier conclusion concerned angles that were much larger than $\gamma$.)  But what matters in these experiments is actually the relative probability $W(\alpha)W(\beta)$ given only the \textit{net} rotation $\delta=\alpha+\beta$ (summing over all the intermediate ways that a given net rotation could occur).  This can be calculated with a straightforward convolution of two Lorentzians:
\begin{equation}
\label{eq:ddef}
W(\delta)= \int^{\infty}_{-\infty} W(\delta+\phi) W(\delta-\phi) d\phi = \frac{a(\gamma)}{\delta^2+(2\gamma)^2}.
\end{equation} 
The result is that Schulman's single-particle anomaly weight also applies to the net rotation of two particles.  The only difference is that $\gamma$ is effectively doubled; $a(\gamma)$ is an irrelevant constant that vanishes when normalized.  

As before, one must sum over all of the different rotations that end up at the same angle, modulo $2\pi$.  The result is an unnormalized joint probability $J(\delta)$ for a net rotation of an angle $0\le \delta < 2\pi$:
\begin{equation}
\label{eq:Jdef}
J(\delta) = \sum\limits_{n=-\infty}^\infty \dfrac{a(\gamma)}{(\delta+ 2n\pi)^2 + (2\gamma)^2}=\dfrac{b(\gamma)}{\sin^2(\delta/2) + \cos^2(\delta/2)\tanh^2(\gamma)}.
\end{equation}
Again, $b(\gamma)$ is an irrelevant constant.

With this exact result, we can reanalyze the above entanglement scenario for a non-zero $\gamma$.  Recall that the initial preparation forces Alice's particle and Bob's particle to initially have opposite spins.  So if they choose the same setting angle ($\theta=0$), and no anomaly occurs, they will always measure opposite outcomes.  Specifically, $J_{A\bar{B}}$, the (unnormalized) joint probability of the outcomes $A$ and $\thickbar{B}$, will be proportional to $J(0)$.  (And so will $J_{\bar{A}{B}}$; the symmetry of this model only concerns relative angles.)  But for a more general setting angle difference of $\theta$, a net anomaly of $\delta=\theta$ will be required for these same anticorrelated outcomes.  Therefore, dropping constants, 
\begin{equation}
\label{eq:Jopp}
J_{A\bar{B}}=J_{\bar{A}{B}}=J(\theta)=\frac{1}{\sin^2\left({\theta}/{2}\right) + \cos^2\left({\theta}/{2} \right)\tanh^2\left(\gamma\right)}.
\end{equation}

Similarly, for correlated outcomes, a net anomaly of $\delta=|\pi-\theta|$ is generally required.  The joint probabilities for these two pairs of outcomes (both with the same required anomaly) are therefore
\begin{equation}
\label{eq:Jsame}
J_{A{B}}=J_{\bar{A}\bar{B}}=J(\pi-\theta)=\frac{1}{\cos^2\left({\theta}/{2}\right) + \sin^2\left({\theta}/{2} \right)\tanh^2\left(\gamma\right)}.
\end{equation}
Note that because of the finite $\gamma$, even if $\theta=0$, (\ref{eq:Jopp}) will no longer be infinitely larger than (\ref{eq:Jsame}).  So even for the same settings, there is still a non-zero probability for correlated outcomes; this is our first direct indication that a finite $\gamma$ will deviate from Born-rule probabilities.

These joint probabilities form a complete set of all possible outcomes (given the boundary constraints), and so can be normalized by dividing by their sum;
\begin{equation}
\label{eq:Zdef}
Z\equiv{J_{AB} + J_{A\thickbar{B}} + J_{\thickbar{A}B} + J_{\thickbar{A}\thickbar{B}}}
\end{equation}
Performing this calculation ($P=J/Z$), the actual probabilities of each of the four outcomes can be extracted.
\begin{eqnarray}
\label{eq:Probs}
P_{A\bar{B}}=P_{\bar{A}{B}}&=& \frac{\cos^2(\theta/2) + \sin^2(\theta/2) \tanh^2(\gamma)}{2+2\tanh^2(\gamma)},\\
P_{A{B}}=P_{\bar{A}\bar{B}}&=&\frac{\sin^2(\theta/2) + \cos^2(\theta/2) \tanh^2(\gamma)}{2+2\tanh^2(\gamma)}.
\end{eqnarray}

Clearly, the Born rule is violated, but not the no-signaling condition!  Bob has effective control over the setting-angle difference $\theta$, but the marginal probability that Alice measures (say) outcome ${A}$ turns out not to depend on $\theta$:
\begin{equation}
\label{eq:NS}
P_{{A}} = (P_{{A}{B}}+P_{{A}\bar{B}}) = \frac{1}{2}.
\end{equation}
This is exact, even for the finite $\gamma$ that deviates the above probabilities from the Born rule.

\subsection{Deviation from the Tsirelson Bound}

The mathematical necessity of the small finite parameter $\gamma$ gives this model a nice feature that may have other uses (beyond fine-tuning arguments).  While it is easy to come up with models that yield different probabilities than quantum theory, it is not particularly easy to come up with a model that smoothly deviates from these probabilities via some continuous parameter.

One point of possible interest in such a model would be to address questions concerning the ultimate nature of the Tsirelson bound.  This bound concerns the key term in the CHSH inequality \cite{CHSH}:
\begin{equation} 
\label{eq:CHSH}
 S \equiv |E(a_{1}, b_{1}) + E(a_{1}, b_{2}) + E(a_{2}, b_{1}) - E(a_{2}, b_{2})| .
\end{equation}
Here $a_1$ and $a_2$ are two possible angle settings for Alice, and $b_1$ and $b_2$ are two possible angle settings for Bob.  For any given settings, the correlation $E$ is
\begin{equation}
\label{Edef}
E=P_{{A}{B}}+P_{\bar{A}\bar{B}}-P_{\bar{A}{B}}-P_{{A}\bar{B}}.
\end{equation}
This value $S$ is provably bounded by 2 for classical systems (for any possible hidden local instructions, sent with the two particles), but can in principle be as large as 4 without allowing signaling (as in the case of the Popescu-Rohrlich boxes \cite{PRbox}).  

In the quantum mechanical case, however, the upper limit of $S$ is known as the Tsirelson bound with the intermediate value: 
\begin{equation} 
\label{eq:TB}
    S_{QM} \le 2\sqrt{2}.
\end{equation}
There has been much speculation concerning why quantum theory lets us get somewhat beyond the classical limit, but not all the way to the conceptual limit.  

The above model provides a unique tool to see how this limit on $S$ can be smoothly varied, via the parameter $\gamma$; at this stage it is unclear whether it will increase or decease.  To start with, we assumed that the maximum value of $S$ will occur at the same angle settings as in the standard quantum case where $\gamma\to 0$:
\begin{equation} 
    \begin{split} \label{eq:settings}
    a_1  &=  0,      \\
    a_2 &= \pi/2,      \\
    b_1  &= \pi/4,      \\
    b_2 &=  -\pi/4.      \\
    \end{split}
\end{equation}

With these angles, the above equations can be utilized to calculate $S(\gamma)$ in the $\gamma\ne 0$ retrocausal model:
\begin{equation}
   \label{eq:Sres}
        S(\gamma) = 2\sqrt{2} \cdot \left( \frac{1-\tanh^2(\gamma)}{1+\tanh^2(\gamma)}  \right) \equiv 2\sqrt{2} \cdot K(\gamma).
\end{equation}
The function $K(\gamma)$ approaches one as $\gamma\to 0$, recovering the standard (\ref{eq:TB}).  But for any other $\gamma$, $K(\gamma)$ is always less than 1; $S(\gamma)$ therefore never exceeds $S_{QM}$.    

We further ensured that the angles (\ref{eq:settings}) used for the settings did indeed provide the maximum value of $S$.  To achieve this, we performed numerical simulation to calculate $S$ when $a_2$, $b_1$ and $b_2$ had small independent angle deviations from (\ref{eq:settings}) .  After many iterations and full combinations of small angle deviations (with a range of $\pm\ {\pi}/{100}$, divided into 200 steps, and $\gamma = 0.1$), we were unable to find any values for which $S$ exceeded $S(\gamma)$ as given in (\ref{eq:Sres}).  Given that $K(\gamma)\le1$, these retrocausal models therefore deviate from the Tsirelson bound, but always on the lower side; the bound is never violated.

\section{Discussion}

One of the particular benefits of explicit hidden-variable models is that one can examine behavior down at the ontological level, even if that behavior is not evident at the higher-level operational level.  This is particularly useful for questions such as causation-without-signaling, because without signaling, there is no operational-level indication that any causation is happening at all.

In this case, the causation is evident.  Taking the interventionist account of causation (which is the only real option that does not utilize a temporal arrow \cite{Price}), causation can be defined in terms of what \textit{would} be different under a counterfactual intervention.  In this case, the intervention is Alice's (or Bob's) choice of the setting angle.  Comparing counterfactual angles is easy for an explicit hidden variable model; one simply runs the model with various settings; any difference in the resulting parameters is \textit{caused} by Alice's (or Bob's) intervention, according to this account.  

Given this definition, it is simple to see that in the above model Alice can \textit{cause} something to happen on Bob's path.  If Alice chooses the same setting as Bob ($\theta=0$), then no anomaly can possibly occur on his path.  But if Alice chooses a different setting, an anomaly \textit{must} occur somewhere on Bob's path.  (Half of the time, anyway, but this is an essential difference from it never occuring.)  This model does not specify exactly when the anomalous rotation of Bob's particle must finish, but it obviously will be complete before it is measured, such that it will conform to the future boundary condition.  So in this model, Alice can certainly cause things to happen well outside of her light cone, over on Bob's particle before it is measured.

Incidentally, the model also demonstrates that the net effect of Alice's outside-the-lightcone causation is \textit{not} action at a distance.  Indeed, this is a continuous model with all the connections on the particle worldlines, and none of the worldlines are ever spacelike.  The net effect may be that she can cause things to be different, at some region that is spacelike-separated from her, but the mechanism behind this effect can clearly be given a Lorentz-covariant account.

And yet, despite this clear spacelike-causation, there is no spacelike-signaling in this model, even when the parameter $\gamma$ is not forced to zero.  The Wood-Spekkens argument \cite{WS} implies that this model must be fine-tuned to prevent signaling, and evidently the fine-tuning does not occur on the obvious free parameter $\gamma$.  Further consideration is clearly required.

Instead of first calculating the probabilities for each of the four outcomes in Section 3.1, we could have simply gone ahead and computed the marginal probability that Alice measures (say) outcome ${A}$, as in (\ref{eq:NS}).  In terms of the unnormalized joint probabilities $J$, this is easy to do:
\begin{equation}
\label{eq:marg}
P_A = \frac{J_{AB}+J_{A\bar{B}}}{J_{AB} + J_{A\thickbar{B}} + J_{\thickbar{A}B} + J_{\thickbar{A}\thickbar{B}}}.
\end{equation}
But looking at the denominator, it is evidently just twice the numerator because 
\begin{eqnarray}
\label{eq:Js}
J_{AB}&=&J_{\thickbar{A}\thickbar{B}}\, , \\
\label{eq:Ja}
J_{\thickbar{A}B}&=&J_{{A}\thickbar{B}}\, .
\end{eqnarray}
It is the equivalence between these outcome probabilities that is seemingly responsible for the inability of Alice or Bob to signal in this model.

Mathematically, the reason that these probabilities are equal is that Schulman's probabilities only depend on the absolute rotation angle, not the direction.  Mirror-imaging the spin state of both particles leads swaps both Alice's outcomes and Bob's outcomes without changing the relative angles. Equations (\ref{eq:Js}) and (\ref{eq:Ja}) follow directly from this reflection symmetry allowed by the original preparation.

Of course, it is certainly possible to violate this symmetry; even in quantum theory, these equations generally only hold for maximally entangled states.  Unfortunately, Schulman's model is not trivially extendable to partially-entangled states; for these cases, a richer local hidden variable space is required \cite{Bloch}, and the corresponding model is still under development \cite{Wharton16}.  But of course Schulman's model can represent separable states, such as sending a spin-up particle to both Alice and Bob, and in that case (\ref{eq:Js}) and (\ref{eq:Ja}) would generally fail.

In the separable case, the signaling disappears because the probabilities on the two particles are independent (learning about one particle's history does not inform anything about the other particle).  This means the joint probabilities are necessarily factorizable; $J_{AB}=J_A J_B$, etc., and (\ref{eq:marg}) simplifies down to the expected $J_A/(J_A + J_{\bar{A}})$.  Nothing Bob can do would have any impact on Alice's measurement at all, even in this retrocausal model.

But the crucial point is that this model \textit{is indeed fine-tuned}, in the sense that there is a much larger space of related retrocausal models for which Eqns (\ref{eq:Js}) and (\ref{eq:Ja}) do not hold, and some of those other models could yield nonlocal signaling.  For example, if $W(\alpha)$ was not equal to $W(-\alpha)$, then non-local signaling would almost certainly be possible.  The ultimate reason that this model does not allow non-local signaling is therefore the symmetry-motivated assumption $W(\alpha)=W(-\alpha)$.  

This explicit analysis shows that ``fine tuning'' is too broad a critique if it rules out symmetry arguments that restrict the form of the model.  It seems to us that any symmetry-based restriction should be deemed a \textit{reasonable} form of fine-tuning; at least, one that should not necessarily carry connotations of unreasonable model-rigging.  In this case, at least, the relevant symmetry entered in via Schulman's effort to find a fully time-symmetric account of the Born rule as applied to a spin-1/2 state.  Given that symmetry considerations restricted the original single-particle model, it should not be particularly surprising that special cancellations (on the signaling level) are a natural consequence.

With this point laid out explicitly, it sheds some doubt on using fine-tuning arguments to assess generic causal models of fundamental physics where natural symmetries can be exact.  Until a model is in hand, it's difficult to see whether the details that lead to operationally-invisible causation are due to some perfect natural symmetry or an unnatural model-rigging.  

Still, the use of fine-tuning arguments in causal discovery algorithms for complex higher-level systems is not particularly tarnished by this analysis, because symmetries at those scales are \textit{not} generally exact.  (Higher-level symmetries can easily be broken.)  And even in the case of quantum entanglement, one can apply this analysis to cases without explicit models.  If operationally-invisible causation is most naturally explained by hidden symmetries, this tells us something important about the best way to causally explain entanglement.  

\section{Conclusions}

The central example in this paper concerned one specific retrocausal model, where the no-signaling condition turned out to be strictly enforced by a natural symmetry.  This same conclusion can be reached from other retrocausal models in the literature, such as an earlier one proposed by Argaman \cite{Argaman}.  In that model, a similar symmetry results from a natural way to distribute the hidden variables among the initial particles (each allowed hidden option has an equal probability).  

If one takes such fundamental symmetries to be the best explanation of the Wood-Spekkens fine-tuning argument, their argument can be redirected in an effort to find the most \textit{natural} causal explanations of Bell-inequality violations.  The insight provided above is that accounts of causation-without-signaling are far more natural in scenarios that utilize symmetries.  The more symmetries that are present in a given model, the more opportunity those symmetries will have to naturally hide the causal dependencies of that model.  This does not merely apply to cases of (effective) nonlocal causation without \textit{nonlocal} signaling, but also related issues such as retrocausation without \textit{backward in time} signaling.  The conclusion is that if one hopes to naturally restrict signaling without {unnatural} fine-tuning, one should look for models with an abundance of symmetries.  These need not be restricted to the symmetries used above in the case of maximally-entangled states; one obvious addition would be Lorentz symmetry.

Consider the three live options for (spacetime-local) causal explanations of Bell-inequality violations, as presented in the Introduction: models that are superluminal, retrocausal, or superdeterministic.  In the case of the latter, some distant past common cause correlates the hidden parameters in the preparation of the entangled state with the decisions of Alice and Bob as to how to set their measurements.  This scenario is sometimes ridiculed as ``conspiratorial'', because this process must work \textit{no matter how} Alice and Bob come to their measurement setting decision.  Alice could use a lottery machine, and Bob could rely on a mere whim.  

Superdeterministic models, therefore, must continue to yield the same correlations even as the relevant system is expanded to include the (different) ways in which Alice and Bob might make their decisions.  To the extent that this bothers most people as ``conspiratorial'', perhaps this underlies a more quantifiable concern: the Wood-Spekkens fine-tuning argument is now in force between systems with no underlying relational symmetry.  Somehow, no matter how different a process Alice and Bob use to make their decisions, these decisions are just correlated enough to yield the right entanglement correlations, and not more.  Consider how simple it would be, given such superdeterministic causal pathways, to literally correlate Alice's and Bob's \textit{decisions}, so that they could signal to each other without any entangled particles at all.  They could merely make separate decisions and know that their choices were correlated because of superdeterministic causes.  Preventing such signaling (while maintaining the perfect non-signaling correlations) would indeed be unnaturally conspiratorial, and it is therefore reasonable to use the Wood-Spekkens argument to rule out any such superdeterminstic causal connection.

Similarly, models with superluminal influence suffer not only from a broken Lorentz symmetry, but also from a special causal role that must be granted to either Alice or Bob, but not both.  In order to preserve some semblance of causal order, such models usually assume there is some (hidden) preferred-reference frame in which these new influences are instantaneous.  Unless Alice and Bob's measurements are exactly simultaneous in this special frame, one of them is always the \textit{cause} of the superluminal influence and the other is the mere \textit{effect}, even though the measured correlations do not indicate that one of them is playing a special role.  If this broken symmetry is restored, where influence is always allowed to go both ways (no matter when the measurements occur), then these models become retrocausal as \textit{well} as superluminal.

Instead, the maximal symmetries seem to lie in the sector of retrocausal models, where the future measurement constraint on the entangled particles is a contributing cause of the earlier hidden variables.  Lorentz symmetry is evidently recoverable; the influence is all on the time-like trajectories, and no matter who ``measures first'', the very same underlying account is always available.  So long as the space of allowed final outcomes provide the constraint, Alice and Bob could use different types of devices to measure the same outcome without breaking this symmetry on the particles themselves.  Furthermore, many of the motivations for considering retrocausal models are based on time-symmetry \cite{Wharton14, WMP, Price, PriceWharton, Evans}, and this would seem to be another resource for ``natural fine-tuning''.  By implying hidden symmetries, the no-signaling aspects of entanglement correlations might themselves be another reason to seriously consider retrocausal models.

The model discussed in this paper is the most explicit spacetime-local and continuous account of entanglement correlations that has yet been developed, and many of its key assumptions may have a deeper and more general explanation \cite{LOQT}.  Despite its inability to address partially-entangled states, it is a clear stepping-stone on a generally unexplored path to a full (retro-)causal explanation of entanglement phenomena.  And when the role of symmetries is properly taken into account, the Wood-Spekkens fine-tuning argument might actually steer us \textit{towards} this particular path, at least as compared to every other spacetime-local option.  The alternative is to prematurely retreat from causal or local explanations entirely, a step that we do not think should be taken while such promising approaches still remain on the table.

\end{document}